\documentclass[%
reprint,
superscriptaddress,
 amsmath,amssymb,
 aps,
prb,
showkeys
]{revtex4-2}

\usepackage{graphicx}
\usepackage{dcolumn}
\usepackage{bm}
\usepackage{siunitx} 
\usepackage{braket}
\usepackage{float}

\usepackage{xcolor}
\usepackage[normalem]{ulem}
\usepackage{soul}
\usepackage{placeins}
\usepackage{physics}
\usepackage[pagebackref=false]{hyperref}
\begin{document}

\author{H.~Dulisch}
\affiliation{JARA-FIT and 2nd Institute of Physics, RWTH Aachen University, 52074 Aachen, Germany,~EU}%
\affiliation{Peter Gr\"unberg Institute  (PGI-9), Forschungszentrum J\"ulich, 52425 J\"ulich,~Germany,~EU}%

\author{D.~Emmerich}
\affiliation{JARA-FIT and 2nd Institute of Physics, RWTH Aachen University, 52074 Aachen, Germany,~EU}%
\affiliation{Peter Gr\"unberg Institute  (PGI-9), Forschungszentrum J\"ulich, 52425 J\"ulich,~Germany,~EU}%

\author{E.~Icking}
\affiliation{JARA-FIT and 2nd Institute of Physics, RWTH Aachen University, 52074 Aachen, Germany,~EU}%
\affiliation{Peter Gr\"unberg Institute  (PGI-9), Forschungszentrum J\"ulich, 52425 J\"ulich,~Germany,~EU}%

\author{K.~Hecker}
\affiliation{JARA-FIT and 2nd Institute of Physics, RWTH Aachen University, 52074 Aachen, Germany,~EU}%
\affiliation{Peter Gr\"unberg Institute  (PGI-9), Forschungszentrum J\"ulich, 52425 J\"ulich,~Germany,~EU}%

\author{S.~Möller}
\affiliation{JARA-FIT and 2nd Institute of Physics, RWTH Aachen University, 52074 Aachen, Germany,~EU}%
\affiliation{Peter Gr\"unberg Institute  (PGI-9), Forschungszentrum J\"ulich, 52425 J\"ulich,~Germany,~EU}%

\author{L.~Müller}
\affiliation{JARA-FIT and 2nd Institute of Physics, RWTH Aachen University, 52074 Aachen, Germany,~EU}%

\author{K.~Watanabe}
\affiliation{Research Center for Electronic and Optical Materials, National Institute for Materials Science, 1-1 Namiki, Tsukuba 305-0044, Japan}
\author{T.~Taniguchi}
\affiliation{Research Center for Materials Nanoarchitectonics, National Institute for Materials Science,  1-1 Namiki, Tsukuba 305-0044, Japan}%

\author{C.~Volk}
\author{C.~Stampfer}
\email{stampfer@physik.rwth-aachen.de}
\affiliation{JARA-FIT and 2nd Institute of Physics, RWTH Aachen University, 52074 Aachen, Germany,~EU}%
\affiliation{Peter Gr\"unberg Institute  (PGI-9), Forschungszentrum J\"ulich, 52425 J\"ulich,~Germany,~EU}%

\title{Electric field tunable spin-orbit gap in a bilayer graphene/WSe$_{2}$ quantum dot}
\date{\today}

\begin{abstract}
We report on the investigation of proximity-induced spin-orbit coupling (SOC) in a heterostructure of bilayer graphene (BLG) and tungsten diselenide (WSe$_2$). 
A BLG quantum dot (QD) in the few-particle regime acts as a sensitive probe for induced SOC.
Finite bias and magnetotransport spectroscopy measurements reveal a significantly enhanced SOC that decreases with the applied displacement field, distinguishing it from pristine BLG. 
Furthermore, our measurements demonstrate a reduced valley $g$-factor at larger displacement fields, consistent with a weaker lateral confinement of the QD. 
Our findings show evidence of the influence of WSe$_2$ across BLG layers, driven by reduced real-space confinement and increased layer localization of the QD states on the BLG layer distant to the WSe$_2$ at higher displacement fields.
This study demonstrates the electrostatic tunability of the spin-orbit gap in BLG/WSe$_2$ heterostructures, which is especially relevant for the field of spintronics and future spin qubit control in BLG QDs.
\end{abstract}
\maketitle

\section{Introduction}
Bernal stacked bilayer graphene (BLG) is characterized by its low intrinsic spin-orbit coupling (SOC). 
Theory predicts a Kane-Mele type SOC strength of around 25-50\,$\mu$eV~\cite{Boettger2007Mar, Gmitra2009Dec, Konschuh2012Mar} whereas experiments have revealed values in a range of 40-80\,$\mu$eV~\cite{Sichau2019Feb,Banszerus2020May,Banszerus2021Sep,Kurzmann2021Mar,Duprez2024Nov}, mainly considered to be enhanced due to proximity coupling to hBN~\cite{Jing2025Apr}.
The integration of transition metal dichalcogenides (TMDs) in graphene-based heterostructures has enabled an interesting approach for manipulating the SOC in two-dimensional materials.
For example, coupling graphene to a TMD, such as tungsten diselenide (WSe$_2$), has shown promising potential to enhance the intrinsically low SOC in graphene while preserving its exceptionally high carrier mobility~\cite{Gmitra2015oct, Gmitra2016apr, Wang2015Sep, Wang2016oct, Yang2017jul, Voelkl2017Sep, Zihlmann2018Feb, Wakamura2018Mar}. 
Of particular interest are BLG/TMDs heterostructures, thanks to the possibility of tuning the band structure of BLG with an external out-of-plane electric displacement field ($D$-field), which introduces a gate-tunable band-gap~\cite{McCann2013, Slizovskiy2021,Icking2022Nov} and localizes the charge carriers predominantly on one of the graphene layers~\cite{Young2011,Khoo2017Nov, Gmitra2017Oct}. Since only the graphene layer in direct contact to the TMD is expected to exhibit significant proximity-induced SOC, this results in a splitting of either the conduction or valence band depending on the sign, i.e. direction of the $D$-field~\cite{Khoo2017Nov, Gmitra2017Oct}. Furthermore,  BLG/WSe$_2$ heterostructures demonstrate high electronic quality with charge carrier mobilities exceeding $100.000\,$cm$^2$(Vs)$^{-1}$ allowing for ballistic transport~\cite{Voelkl2017Sep,Wang2019Oct} and large spin diffusion lengths~\cite{Drogeler2014Nov,aynes2015nov,Bisswanger2022Jun}, making it suitable for the development of spin transistors~\cite{HH2009Sep,Han2011Jul}. 
This unique combination of high carrier mobility, gate-tunable SOC, and spin-dependent electronic band properties makes BLG/TMD heterostructures a versatile platform for spintronics applications, with implications for the design of next-generation spin-based transistors and quantum devices.

Until now, proximity-induced SOC has been investigated predominantly in bulk-like BLG/WSe$_2$ devices, where the nearly perfect layer localization at low energies keeps band mixing to a minimum. 
However, when lateral confinement is introduced — as in gate-defined QDs in BLG — the dot wavefunction includes contributions from higher-momentum states. This leads to reduced layer localization and a modified SOC strength.
Here, we investigate the proximity effect of WSe\(_{2}\) on the SOC in a BLG hole QD, where the QD wavefunction predominantly occupies the graphene layer opposite to the WSe\(_{2}\) layer. 
By magnetotransport and finite-bias spectroscopy measurements in the single charge-carrier regime, we show that the spin-orbit gap \(\Delta_{\mathrm{SO}}\) in the BLG-QD exhibit a pronounced dependence on the displacement field $D$, increasing at lower $D$-fields, while the spin \(g\)-factor remains largely unchanged. 
Additionally, we find that the valley \(g\)-factor decreases for larger displacement fields, indicating a lateral size increase of the QD. We attribute this trend to the interplay of enhanced layer localization in the BLG conduction and valence bands and a reduction in lateral confinement at increased displacement fields.

\begin{figure}[!thb]  
    \centering
    \includegraphics[draft=False,keepaspectratio=true,clip,width=\linewidth]{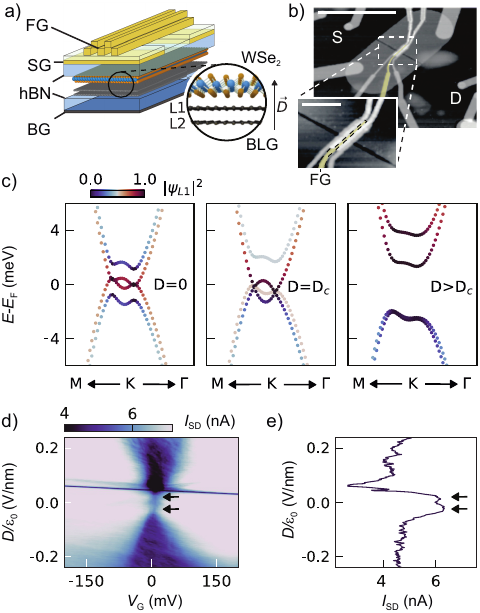}
    \caption{(a) Schematic of the device showing the relevant metal gate structure and the van der Waals heterostructure. The inset highlights the BLG/WSe$_2$ stacking order.
    (b) Scanning force micrograph of the finished device showing the metal gate structure. 
    (c) Tight-binding band structure calculations close to the $K$ point at $D$-fields corresponding to $D=0$, $D=D_\mathrm{c}$ and $D>D_\mathrm{c}$, respectively. The color scale corresponds to the expectation value of the projection operator on the graphene layer closer to the WSe$_2$ (L1). The tight-binding band structure calculations are based on~\cite{Konschuh2012Mar}. In our calculations we employ their model with the parameters listed in table 1 in \cite{Konschuh2012Mar}, with the slight change of adding $\lambda_{\Delta}$ to the parameters $\lambda_{\text{I1}}$ and $\lambda_{\text{I2}}$ to the block-matrix of $\mathcal{H}_{\text{SOC}}$ which corresponds to layer 1 in the onsite orbital basis $\Psi_{\text{A$_i$}}$, $\Psi_{\text{B$_i$}}$, where $i$ denotes the layer index.
    (d) $I_\mathrm{SD}$ through the conducting channel as a function of the effective gate voltage $V_\mathrm{G}$ and $D/\epsilon_0$ with no voltage applied to the finger gates. The feature close to zero displacement field and zero doping is attributed to the band inverted phase appearing in TMD/BLG heterostructures due to increased Ising-type SOC, which leads to a dip in conductance between the two critical field values (black arrows) $\pm D_\mathrm{c}/\epsilon_0$~\cite{seiler_layer-selective_2024,Masseroni2024Oct}. 
    (e) Line cut at $V_\text{G} = 0$ (see panel (d)) showing the decrease in current close to $D$/$\epsilon_0=0$. 
    }
    \label{f0}
\end{figure}

\section{Device fabrication and operation}
The device consists of a monolayer of WSe$_2$ on top of a BLG flake, both encapsulated between two approximately 40~nm thick crystals of hexagonal boron nitride (hBN).
This van der Waals heterostructure is placed on top of a graphite flake that functions as a back gate (BG). 
Three metal (Cr/Au) gate layers -- split gates (SGs) and two layers of interdigitated finger gates (FGs), -- are deposited on top of the heterostructure, each separated by a 20~nm thick layer of Al$_2$O$_3$. 
Fig.~\ref{f0}(a) shows a schematic of the heterostructure, whereas Fig.~\ref{f0}(b) shows a scanning force microscope image of the gate structure.
All the experiments were performed in a dilution refrigerator at a base temperature of around 30\,mK.
\begin{figure}[!thb]  
    \centering
    \includegraphics[draft=False,keepaspectratio=true,clip,width=\linewidth]{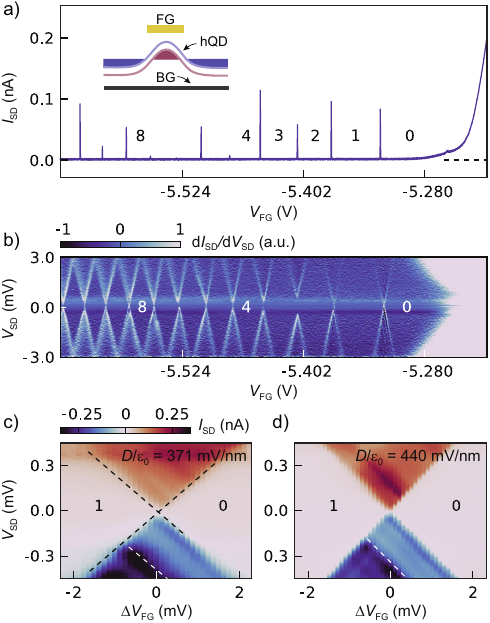}
    \caption{(a) $I_\mathrm{SD}$ as a function of $V_\mathrm{FG}$ showing the first few Coulomb peaks of a hole QD measured at a displacement field of $D/\epsilon_0 = 440$\,mV/nm and $B_\perp = 200$\,mT. The inset shows a schematic of the band-edge profile along the channel, with the QD being created by the induced n-p-n junction due to the local inversion of the bands under the FG.
    (b) Finite-bias spectroscopy showing the differential conductance d$I$/d$V$ at $D/\epsilon_0 = 440$\,mV/nm and $B_\perp = 0$\,T.
    (c) and (d) show a close up on the first charge transition at $D/\epsilon_0 = 371$\,mV/nm (c) and $D/\epsilon_0 = 440$\,mV/nm (d) respectively at $B_\perp = 0$\,T. The white dashed line indicates the first excited state and its intercept with the outline of the conductive region.}
    \label{f1}
\end{figure}

The SGs are used to open a band gap in the BLG beneath them, creating a narrow ($\sim$ 200~nm wide) conducting channel connecting the source (S) and drain (D) contacts.
The voltages applied to the BG and SGs allow to independently tune the effective gate voltage, $V_\mathrm{G} = (V_\mathrm{BG}-V_\mathrm{BG}^0 + \beta(V_\mathrm{SG}-V_\mathrm{SG}^0))/(1+\beta)$, and the displacement field, $D/\epsilon_{0} = e/2 \cdot(\alpha_\mathrm{BG}(V_\mathrm{BG}-V_\mathrm{BG}^0)-\alpha_\mathrm{SG}(V_\mathrm{SG}-V_\mathrm{SG}^0))$, where $\beta=\alpha_\mathrm{BG}/\alpha_\mathrm{SG}$ is the ratio between the geometric lever arms of the BG and the SGs respectively, while $V^{0}_\mathrm{BG}$ and $V^{0}_\mathrm{SG}$ are the offset voltages of the charge neutrality point~\cite{Icking2022Nov}.
Figure~\ref{f0}(d) shows the source-drain current $I_\mathrm{SD}$ as a function of $V_\mathrm{G}$ and $D/\epsilon_0$. 
The data shows a non-monotonic dependency of the current near the charge neutrality point at low displacement fields. 
A cut through the current map at $V_\mathrm{G}=0$\,V reveals a local minimum in $I_\mathrm{SD}$ at $D/\epsilon_0=0$\,mV/nm and two symmetrically centered local maxima at the critical field $D = \pm D_\mathrm{c}$ (indicated by the black arrows in Figs.~\ref{f0}(d) and (e)). 
Similar features as in Fig.~\ref{f0}(d) have been observed in conductance measurements of dual gated suspended BLG devices, which was attributed to a symmetry broken state due to electron-electron interaction~\cite{weitz_broken-symmetry_2010}.
Recently, this inverted gap phase was also observed in transport measurements of BLG fully encapsulated between WSe$_2$~\cite{Island2019Jun}, as well as in samples with an one-sided contact to TMDs, such as WSe$_2$ \cite{seiler_layer-selective_2024} and MoS$_2$ \cite{Masseroni2024Oct}.  
In this case, the feature is attributed to band inversion caused by induced Ising-type SOC. Indeed, tight-binding band-structure calculations indicate the existence of a critical field $D_\mathrm{c}$ for which the SOC-induced gap closes, see Fig.~\ref{f0}(c). At this critical field, the band gap closes showing up as peaks in the current (c.f. Fig.~\ref{f0}(e)), while the breaking of layer symmetry leads to a dip in current at $D/\epsilon_0=0$\,V/nm. 
The signature of this inverted gap phase is less pronounced in our data, due to our signal being dominated  by the induced conductive channel in the device, which is only gated by the BG.   
Nevertheless, this is a clear evidence of an enhanced proximity-induces spin-orbit coupling in the BLG.

The finger-gate structure on top of the split-gate allows us to modulate the potential profile along the conducting channel and to form a quantum dot, see inset in Fig.~\ref{f1}(a)~\cite{banszerus2018gate,eich2018spin,banszerus2020electron}. 
We deplete the channel to form an electrostatically defined QD in BLG, by applying a voltage $\text{V}_\mathrm{FG}$ to the central finger gate, while the remaining FGs are used to tune the tunnel barriers and isolate the QD.
Figure~\ref{f1}(a) shows the complete pinch-off of the current through the channel for increasing $\text{V}_\mathrm{FG}$, and the appearance of well-defined Coulomb peaks indicating the sequential filling of a  QD with holes.
Figure~\ref{f1}(b) shows a finite-bias spectroscopy measurement, i.e. the normalized differential conductance as a function of $\text{V}_\mathrm{FG}$ and the bias voltage $V_\mathrm{SD}$. 
From the characteristic Coulomb diamond signature, we extract the gate lever arm $\alpha = \Delta \text{V}_\mathrm{SD}/\Delta \text{V}_\mathrm{FG}$, which allows us to translate changes in gate voltage $\text{V}_\mathrm{FG}$ into changes in electrochemical potential $\Delta \mu$. 

\begin{figure}[!thb]  
    \centering
    \includegraphics[draft=False,keepaspectratio=true,clip,width=\linewidth]{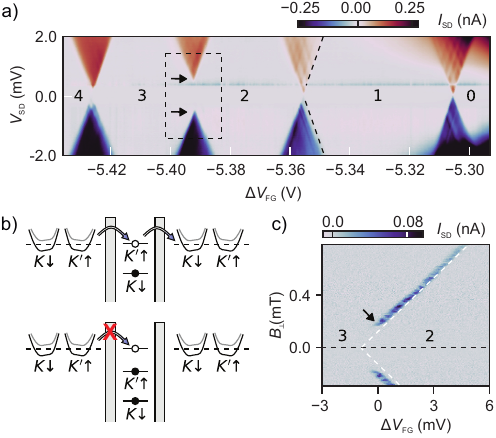}
    \caption{(a) Finite-bias spectroscopy of the first 4 holes in the QD measured at $B=0$\,T.
    (b) Schematics depicting sequential tunneling through a LG QD at the 1 to 2 transition (upper schematic) and the 2 to 3 transition (lower schematic) with spin-valley split bands in the source-drain lead regions (for more details see text). 
    (c) Magneto transport spectroscopy measurement of the $3^\text{rd}$ hole at $V_\mathrm{SD} = 60\,\mu$V highlighting that transport only becomes possible at finite magnetic field. 
   }
    \label{f3N}
\end{figure}

The Coulomb-blockade measurements show also signatures of the presence of a proximity-induced SOC gap in the BLG. 
Indeed, we observe that the third Coulomb diamond is not properly closing (see dashed box in Fig.~\ref{f3N}(a)). 
This can be understood in terms of an imbalance of the four flavors ($\ket{K\uparrow}$, $\ket{K\downarrow}$, $\ket{K'\uparrow}$ and $\ket{K'\downarrow}$) in the lead regions, which for large displacement field, have a band-structure with spin split bands as shown as in the right panel of Fig.~\ref{f0}c.
Due to this imbalance, the first two holes, $\ket{K'\uparrow}$ and $\ket{K\downarrow}$ can be easily filled and low-bias transport is possible (see upper schematic of Fig.~\ref{f3N}(b)).
However, the process of adding the $3^\textbf{rd}$ hole, which corresponds to a $\ket{K'\downarrow}$ or $\ket{K\uparrow}$ state, is suppressed (see lower schematic).
The blockade can be either lifted by the bias voltage (see arrows in Fig.~\ref{f3N}(a))  making it possible to populate $\ket{K'\downarrow}$ or $\ket{K\uparrow}$ states in the leads or by an out-of-plane $B$-field (see Fig.~\ref{f3N}(c)).
The latter will shift the states of the QD according to their spin and valley $g$-factors, changing the state ordering with increasing $B_\perp$ (for details see Ref.~\cite{Moller2023Sep}).
In particular, a state of the $2^\textbf{nd}$ orbital, decreasing in energy, can become the new three-particle ground state, making transport possible.

Finite-bias spectroscopy measurements also allow to investigate the excited states of the QD \cite{kurzmann_excited_2019}, as shown  in Figs.~\ref{f1}(c) and (d) for two different displacement field. This type of measurement allows to determine the single-particle level splitting. 
We assume the first excited state to coincide with the SO-gap. From finite-bias spectroscopy measurements we estimate an orbital splitting of $\sim500\,\mu$eV.
As discussed below, the values extracted for the first excited state agrees very well with the SOC strength estimated by magnetic field spectroscopy. The data of Fig.~\ref{f1}c present therefore a first indication that the SOC in the dot depends on the applied displacement field.  
\begin{figure}[!thb]  
    \centering
    \includegraphics[draft=False,keepaspectratio=true,clip,width=\linewidth]{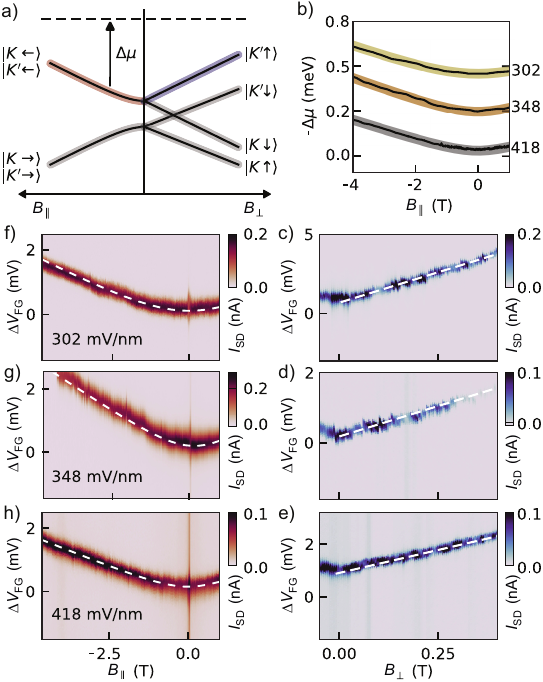}
    \caption{
    (a) Schematic of the single particle energy levels in dependence of the applied magnetic field. The black arrow indicates the difference in electrochemical potential of the dot level with respect to the leads. 
    (b) Three exemplary datasets extracted from the in-plane magneto-transport data  shown in panel (f)-(h) together with the fits according to Eq.~(2). The datasets are offset for better visibility. 
    (c)-(e) Out-of-plane magneto-transport spectroscopy measurements for (c) $D/\epsilon_0=$ 302 mV/nm, (d) $D/\epsilon_0=$ 348 mV/nm, (e) $D/\epsilon_0=$ 418 mV/nm. 
    The fits according to Eq.~(1) are overlaid as a dashed line for comparison.
    (f)-(h) In-plane magneto transport data for the same displacement fields as panel (c)-(e).
   }
    \label{f2}
\end{figure}

\section{Magnetic field spectroscopy}
The size of the SOC gap in the dot, $\Delta_\mathrm{SO}$, and the values of the spin and valley $g$-factors, $g_\mathrm{s}$ and $g_\mathrm{v}$ can be experimentally determined by measuring the shift of the first Coulomb resonance as function of an external magnetic field, since this reflect directly the change in ground-state energy of the single particle states, see Fig.~\ref{f2}(a).
This exhibits a linear dependence on the out-of-plane magnetic field $B_{\perp}$ given by the relation
\begin{equation}
    \Delta \mu = \frac{1}{2} (g_\mathrm{s} + g_\mathrm{v}) \mu_\mathrm{B} B_{\perp},
\end{equation}
and the following dependence on the in-plane field $B_{\parallel}$:
\begin{equation}
\Delta \mu = \frac{1}{2}\sqrt{\Delta_\mathrm{SO}^{2} + (g_\mathrm{s}\mu_\mathrm{B}B_{\parallel})^{2}}
\label{eq1}
\end{equation}
where $\mu_\mathrm{B}$ is the Bohr magneton, $g_\mathrm{s}$ the spin $g$-factor, $g_\mathrm{v}$ is the valley $g$-factor, originating from the orbital valley magnetic moment caused by a non-vanishing Berry curvature at the high symmetry points $K$ and $K'$~\cite{Knothe2020Jun}, and $\Delta_\mathrm{SO}$ is the spin-orbit gap. 
The shift of the first Coulomb resonance as function of $B_{\parallel}$ and $B_\perp$ are shown in Figs.~\ref{f2}(c)-\ref{f2}(h) for different values of the applied displacement field $D$. 
Knowing the gate lever arm $\alpha$ from finite-bias spectroscopy measurements, it is possible to translate the shift of the resonance in gate-voltage $V_\mathrm{FG}$ into a change of electrochemical potential $\Delta\mu=\alpha \Delta V_\mathrm{FG}$, as shown in Fig.~4(b) for the measurements as function of $B_\parallel$. 

\begin{figure}[!thb]  
    \centering
    \includegraphics[draft=False,keepaspectratio=true,clip,width=\linewidth]{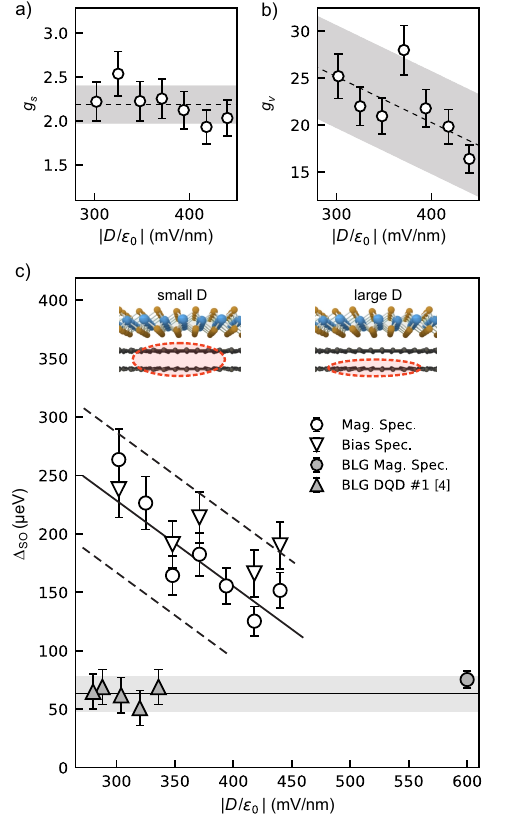}
  \caption{
    (a) Spin $g$-factor $g_\mathrm{s}$ determined from in-plane magnetic field spectroscopy measurements for different displacement field strengths $|D/ \epsilon_0|$. The data scatter around a mean value of $2.2 \pm 0.2$, indicated by the filled area, with no discernible trend. 
    (b) Same as (a), but for the out-of-plane field direction, showing the valley-g factor obtained by subtracting the spin $g$-factor from the slope of a linear fit to the data. The valley $g$-factor decreases with increasing electric field ($D$-field), suggesting reduced lateral confinement of the QD, consistent with \cite{Moller2023Sep}.
    (c) Comparison of $\Delta_\mathrm{SO}$ as a function of $|D/\epsilon_0|$ for BLG/WSe$_{2}$ (white) and pure BLG QD (grey) devices. The two methods used to extract the SOC parameter in the case of the BLG/WSe$_{2}$ device show good agreement and a decreasing trend with increasing displacement field strength, absent in the pure BLG data. The data for the pristine BLG QD devices stems from ref.~\cite{Banszerus2021Sep}.
}
    \label{f3}
\end{figure}

At large magnetic field, the peak position shifts approximately linearly with $B_\parallel$, $
\Delta \mu \approx \frac{1}{2} g_\mathrm{s}\mu_\mathrm{B} B_{\parallel} $, which allows to extract the spin $g$-factor independently.
As expected, the values of $g_\mathrm{s}$ extracted in this way show no dependence on $|D/\epsilon_0|$ within the margin of error, see Fig.~\ref{f3}(a). 
The main source of uncertainty arises from the determination of the lever arm, which depends on identifying the onset of the conducting region in the finite-bias spectroscopy measurements. Due to the inherent ambiguity in this onset, we estimate a systematic error of approximately~$\sim10\%$.
Taking the mean value of $g_\mathrm{s}$ determined in this way as fixed parameter, $g_\mathrm{s}=2.2\pm 0.2$, we fit Eq.~(1) and Eq.~(2) to the data of Fig.~\ref{f2}, to determine $g_\mathrm{v}$ and $\Delta_\mathrm{SO}$ respectively.  The values determined in this way are summarized in Fig.~\ref{f3}. In Fig.~5(c) we include also values of $\Delta_\mathrm{SO}$ determined with finite-bias spectroscopy (down-pointing white triangles), which agree very well with those extracted with the procedure discussed above. 

Differently from the spin $g$-factor, both the valley $g$-factor, $g_\mathrm{v}$, and the spin-orbit coupling $\Delta_\mathrm{SO}$ show a clear dependence on the applied displacement field, both decreasing with increasing $|D/\varepsilon_0|$. The decrease of $g_\mathrm{v}$ indicates a weaker lateral confinement for larger $|D/\varepsilon_0|$, in agreement with earlier work~\cite{Moller2023Sep}. 
This weaker lateral confinement is also one of the effects that leads to the observed reduction of $\Delta_\mathrm{SO}$ for increasing $|D/\varepsilon_0|$, as discussed in the following section. 

\section{Tunable proximity-induced spin-orbit gap}
At low displacement field, the extracted SOC strength is significantly larger than what was observed in QDs formed in pure BLG-hBN heterostructures~\cite{Banszerus2021Sep,Banszerus2023Jun}, see also gray data-points in Fig.~\ref{f3}(c)). Such an enhancement is a clear indication of proximity-induced SOC caused by the WSe\(_2\) layer in our heterostructure.
Interestingly, our QD is expected to be located on the BLG layer opposite to the WSe\(_2\). 
In fact, in gapped BLG, the valence and conduction bands near the \(K\)- and \(K'\)-valleys are layer-polarized: one layer predominantly hosts the conduction band, while the other hosts the valence band. 
The polarity of the displacement field determines which band localizes on the top (WSe\(_2\)-side) or bottom graphene layer~\cite{Khoo2017Nov, Gmitra2017Oct}. 
In our device, the conduction band is localized on the top-layer, which implies that for negative doping, we fill states in the valence band localized on the bottom graphene layer. 
To form a QD in the hole regime, the charge-carrier polarity must be locally inverted, so the QD wavefunction predominantly derives from bottom-layer states.
Despite this nominal separation, the measurements reveal a substantial proximity-induced SOC effect. 

One reason for this is that the layer localization is most pronounced right at the \(K\) and \(K'\)-points, while  
away from these high-symmetry points, the Bloch wavefunctions experience increased mixing across both layers, providing overlap with the top graphene layer.  In the quantum dot, the real-space confinement of the wavefunction results in momentum-space mixing of Bloch states, enhancing the contribution from the proximitized BLG layer and resulting in the observed enhanced $\Delta_\mathrm{SO}$.  
However, for increasing the displacement field, the layer-polarization (i.e. layer localization) of the wavefunction increases (see schematics in Fig.~5(c)) and, at the same time, the lateral confinement of the dot is reduced~\cite{Moller2023Sep}, as indicated by decreasing values of the valley $g$-factor, see Fig.~\ref{f3}(b), narrowing the distribution of the wave function in momentum space around layer-polarized states. The two effects contributes to the observed decrease of $\Delta_\mathrm{SO}$ with increasing $|D/\varepsilon_0|$. 

\section{Conclusion}
In summary, we have experimentally investigated the proximity-induced SOC in a BLG QD formed adjacent to a WSe\(_{2}\) layer.
Through magnetotransport and finite-bias spectroscopy measurements in the few-carrier regime, we find a pronounced enhancement of the SOC gap \(\Delta_{\mathrm{SO}}\) compared to well studied pure BLG QDs, underscoring the impact of the TMD on nearby graphene layers.
This spin-orbit gap decreases with increasing \(|D/\epsilon_0|\), which is in contrast to pristine BLG QDs, where no tuning of \(\Delta_{\mathrm{SO}}\) with \(D/\epsilon_0\) was observed.
This behavior can be understood as the interplay of two key effects: (i) a larger band gap strengthens layer localization at the high-symmetry points \(K,\,K'\), and (ii) the lateral confinement of the QD decreases at higher \(|D/\epsilon_0|\), causing the wavefunction to include fewer mixed-layer states that would otherwise enhance the SOC strength.

Notably, we observe that the QD primarily resides on the graphene layer opposite to the WSe\(_{2}\), yet still exhibits a sizable proximity-induced SOC.
This highlights the importance of momentum-space mixing of Bloch states and shows that a purely layer-localized picture is insufficient at experimentally relevant gate voltages and QD sizes.
Furthermore, our out-of-plane magnetotransport data reveal a systematic reduction of the valley \(g\)-factor with increasing \(|D/\epsilon_0|\), consistent with reduced lateral confinement and a larger real-space extent of the QD.
Altogether, these findings demonstrate robust, tunable SOC in a BLG/WSe\(_{2}\) heterostructure, paving the way for future spintronic and QD-based spin qubit experiments where electrical control of spin-orbit coupling may be used to realize spin-orbit driven qubits and spin-based logic devices.\\

While finalizing the manuscript we became aware of another recent work studying a BLG QD in proximity to a WSe$_2$ layer~\cite{Gerber2025Apr}. Interestingly, their observed dependency of the proximity enhanced spin orbit gap on \(|D/\epsilon_0|\) is in very good qualitative and quantitative agreement supporting our findings.\\
\newline
\textbf{Acknowledgments  }
The authors thank F.~Lentz, S.~Trellenkamp and M.~Otto for help with sample fabrication and F. Haupt for help on the manuscript.
This project has received funding from the European Research Council (ERC) under grant agreement No. 820254, the Deutsche Forschungsgemeinschaft (DFG, German Research Foundation) through SPP 2244 (Project No. 535377524) and under Germany’s Excellence Strategy - Cluster of Excellence Matter and Light for Quantum Computing (ML4Q) EXC 2004/1-390534769, and by the Helmholtz Nano Facility~\cite{Albrecht2017May}. 
K.W. and T.T. acknowledge support from the JSPS KAKENHI (Grant Numbers 21H05233 and 23H02052) , the CREST (JPMJCR24A5), JST and World Premier International Research Center Initiative (WPI), MEXT, Japan.\\

\textbf{Author contributions  }
  H.D, D.E. and E.I. fabricated the device. H.D. and D.E. performed the measurements and analyzed the data with the help of L.M. and S.M. K.W. and T.T.  synthesized the hBN crystals. C.V. and C.S. supervised the project. H.D., D.E., K.H., C.V., and C.S. wrote the manuscript with contributions from all authors.\\

\textbf{Data availability}
The data underlying this study are openly available in a Zenodo repository at DOI XXX.

\textbf{Competing interests  }
The authors declare no competing interests.


\providecommand{\latin}[1]{#1}
\makeatletter
\providecommand{\doi}
  {\begingroup\let\do\@makeother\dospecials
  \catcode`\{=1 \catcode`\}=2 \doi@aux}
\providecommand{\doi@aux}[1]{\endgroup\texttt{#1}}
\makeatother
\providecommand*\mcitethebibliography{\thebibliography}
\csname @ifundefined\endcsname{endmcitethebibliography}  {\let\endmcitethebibliography\endthebibliography}{}

\end{document}